\documentclass{aastex631}

\defcitealias{original}{C21}

\begin{document}

\title{Caught in the Act: Observations of the Double-mode 
RR Lyrae\\V338 Boo During the Disappearance of a Pulsation Mode}

\correspondingauthor{Kenneth Carrell}
\email{kenneth.carrell@angelo.edu}

\author[0000-0002-6307-992X]{Kenneth Carrell}
\affiliation{Department of Physics \& Geosciences, Angelo State University, 
San Angelo, TX 76909, USA}

\author[0000-0002-4792-7722]{Ronald Wilhelm}
\affiliation{Department of Physics \& Astronomy, University 
of Kentucky, Lexington, KY 40506, USA}

\author{Andrew Tom}
\affiliation{Department of Physics \& Geosciences, Angelo State University, 
San Angelo, TX 76909, USA}

\author{Horace Smith}
\affiliation{Department of Physics \& Astronomy, Michigan State University, East 
Lansing, MI 48824, USA}

\author[0000-0003-3184-5228]{Adam Popowicz}
\affiliation{obscode: PDM, Department of Electronics, Electrical Engineering, and 
Microelectronics, Silesian University of Technology, 44-100 Gliwice, Poland}

\author{Gary Hug}
\affiliation{obscode: HGAG, Sandlot Observatory MPC H36, Scranton, KS 66537, USA}

\author[0000-0002-9205-5329]{Stephen M.~Brincat}
\affiliation{obscode: BSM, Flarestar Observatory, San Gwann, Malta}

\author{Fabio Salvaggio}
\affiliation{obscode: SFV, Wild Boar Remote Observatory (K49), Florence, Italy}

\author{Keith Nakonechny}
\affiliation{obscode: NKEA, Barrie, Ontario, Canada}

\author{Darrell Lee}
\affiliation{obscode: LDRB, Friendswood, TX, USA}

\author{Te\'{o}filo Arranz Heras}
\affiliation{obscode: ATE, Las Pegueras Observatory, Navas de Oro, Segovia, Spain}

\author{Tony Vale}
\affiliation{obscode: VTY, Bowerhill Observatory, Melksham, Wiltshire, UK}

\author{Davide Mortari}
\affiliation{obscode: MFAA}

\author{Andr\'{e} Steenkamp}
\affiliation{obscode: SABB, Southwater Observatory, Southwater, Horsham, West Sussex, UK}

\author{Ralph Rogge}
\affiliation{obscode: RRO, Constance, Germany}

\author[0000-0001-6352-730X]{Jacek Checinski}
\affiliation{Department of Electronics, Electrical Engineering, and 
Microelectronics, Silesian University of Technology, 44-100 Gliwice, Poland}

\begin{abstract}
New results on the behavior of the double-mode RR Lyrae V338 Boo are presented. 
The Transiting Exoplanet Survey Satellite (TESS) observed this star again in 2022, 
and an observing campaign of the American Association of Variable Star Observers 
(AAVSO) was completed after the TESS observations as a follow-up. We find that 
the first overtone pulsation mode in this star completely disappears during the 
TESS observing window. This mode reappears at the end of the TESS observations, 
and the AAVSO observing campaign shows that in the months that followed, the 
first overtone mode was not only present, but was the dominant mode of 
pulsation. This star, and potentially others like it, could hold the key to finally 
solving the mystery of the Blazhko effect in RR Lyrae.
\end{abstract}

\section{Introduction}\label{sec:intro}
\subsection{Background}\label{ssec:background}
Pioneering work in the late 19th century using the new technique of 
photography led to the discovery of a bright (seventh magnitude) 
variable star in the constellation of Lyra by Williamina Fleming 
\citep{RRLdiscovery} that was designated RR Lyrae. This star became 
the archetype of a distinct class of variable star identified in old stellar 
populations. \citet{bailey} sub-classified RR Lyrae into three types 
({\it a}, {\it b}, and {\it c}) that were simplified into two types 
that are still in use today: {\it ab} (RRab) and {\it c} (RRc). These are now 
believed to be evolved He-core burning, low-mass stars in the instability strip 
of the HR Diagram with radial pulsations in the fundamental (RRab) 
and first overtone (RRc) modes \citep[more detail, and additional 
references, can be found in][]{book,book2}. It was noted early on that some 
RRc types had a scatter not explainable by observational error. It was 
not until much later, however, that changes in the light curve due to 
mode mixing of the fundamental and first overtone radial pulsation 
modes were found in AQ Leo \citep{AQLeo} and RR Lyrae in the 
globular cluster M15 \citep{RRdM15}. This sub-class of 
double-mode pulsators, which simultaneously pulsate in both the 
fundamental and first overtone modes, is classified as type {\it d} 
(RRd). The comprehensive review of pulsation theory up until that 
point by \citet{christy-theory} set the stage for \citet{jorgpet} to 
recognize that double-mode pulsations could be important 
probes of physical properties of these stars. Then \citet{petersendiagram} 
introduced the diagram (later named after him) that allows a mass 
estimate of double-mode pulsators. RRd stars in particular were 
used to test differences between stellar evolution and stellar 
pulsation models using opacities in stellar interiors 
\citep{rrd-opal} and to improve and verify nonlinear stellar pulsation 
models that include time-dependent turbulent convection 
\citep{nonlinearmodel}. More recently, luminosities of RRd stars 
from models and observations were compared \citep{rrd-gaia}, 
which may help resolve the current so-called ``Hubble 
tension'' by using Population II stars \citep{popIIdist}.

Other multi-periodic behavior has been identified among the RR 
Lyrae variables. The Blazhko effect, a periodic modulation of the 
primary light cycle, was identified in RRab stars by \citet{blazhko} 
and \citet{shapley}, and was later identified in RRc type stars. The 
Blazhko period is often tens of days long, and the physical mechanism 
causing this behavior remains uncertain, making the Blazhko effect one 
of the longest unanswered questions in all of astrophysics. With the 
unique insight available with RRd type stars, however, any of these 
variables exhibiting the Blazhko effect are much more interesting 
and important, and could hold the key to solving this more than 
century-old puzzle.

\subsection{V338 Boo}\label{ssec:introv338}
The RRd star V338 Boo has been studied for almost two decades 
now, and with each set of observations there are new and 
interesting results.

The work of \citet{oaster06} showed that the period ratio of the first overtone 
to fundamental modes was normal for an RRd ($P_1/P_0$ = 0.743) but 
the fundamental mode amplitude was twice as large as the first overtone. 
Follow-up observations that were published four \citep{jaavso} and 
six \citep{jaavso2} years later showed that 
the amplitude ratio of the two pulsation modes was changing over time.

In our original paper \citep{original}, hereafter \citetalias{original}, we 
presented Transiting Exoplanet Survey Satellite \citep[TESS,][]{tess} 
observations of V338 Boo from 2020 (Sectors 23 and 24). These 
observations showed that this star was actually undergoing changes in 
the amplitudes of its pulsation modes over the course of tens of days,
and we were able to explain what was seen from earlier, ground-based, 
data. The survey strategy of TESS is such that most targets are re-observed 
every other year, which was the case for V338 Boo - it was observed again 
by TESS in Sectors 50 and 51, which correspond to 2022 March 26 to April 
22 and 2022 April 22 to May 18, respectively.

Because we knew this star would be observed again by TESS in 2022, 
we were able to plan a campaign through the American Association of 
Variable Star Observers (AAVSO) to extend the baseline of observations 
and verify what was seen in TESS data.

In the following, we will describe the newer TESS observations 
(Section \ref{ssec:tessdata}), the follow-up observations by AAVSO 
observers (Section \ref{ssec:aavsodata}), and new behavior 
for V338 Boo (Section \ref{sec:results}). In the end we will 
describe our conclusions and discuss what they might mean 
(Section \ref{sec:conclusions}).

\section{Data}\label{sec:data}
\subsection{TESS Observations in Sectors 50 and 51}\label{ssec:tessdata}
In \citetalias{original} we presented results from TESS Sectors 
23 and 24, which occurred from 2020 March 18 to May 13. The results 
from this timeframe showed that V338 Boo started as a normal RRd 
star, with the first overtone mode having a higher amplitude than the 
fundamental mode. Over the course of those observations, however, 
the fundamental mode grew by a factor of 4 to 5, becoming the 
dominant mode in the later part. The first overtone mode decreased 
slightly, changing by less than a factor of 2. What was unknown in 
\citetalias{original} was how this star behaved after this 
observational window.

\begin{figure*}[htb]
  \centering
  \includegraphics[width=0.45\textwidth]{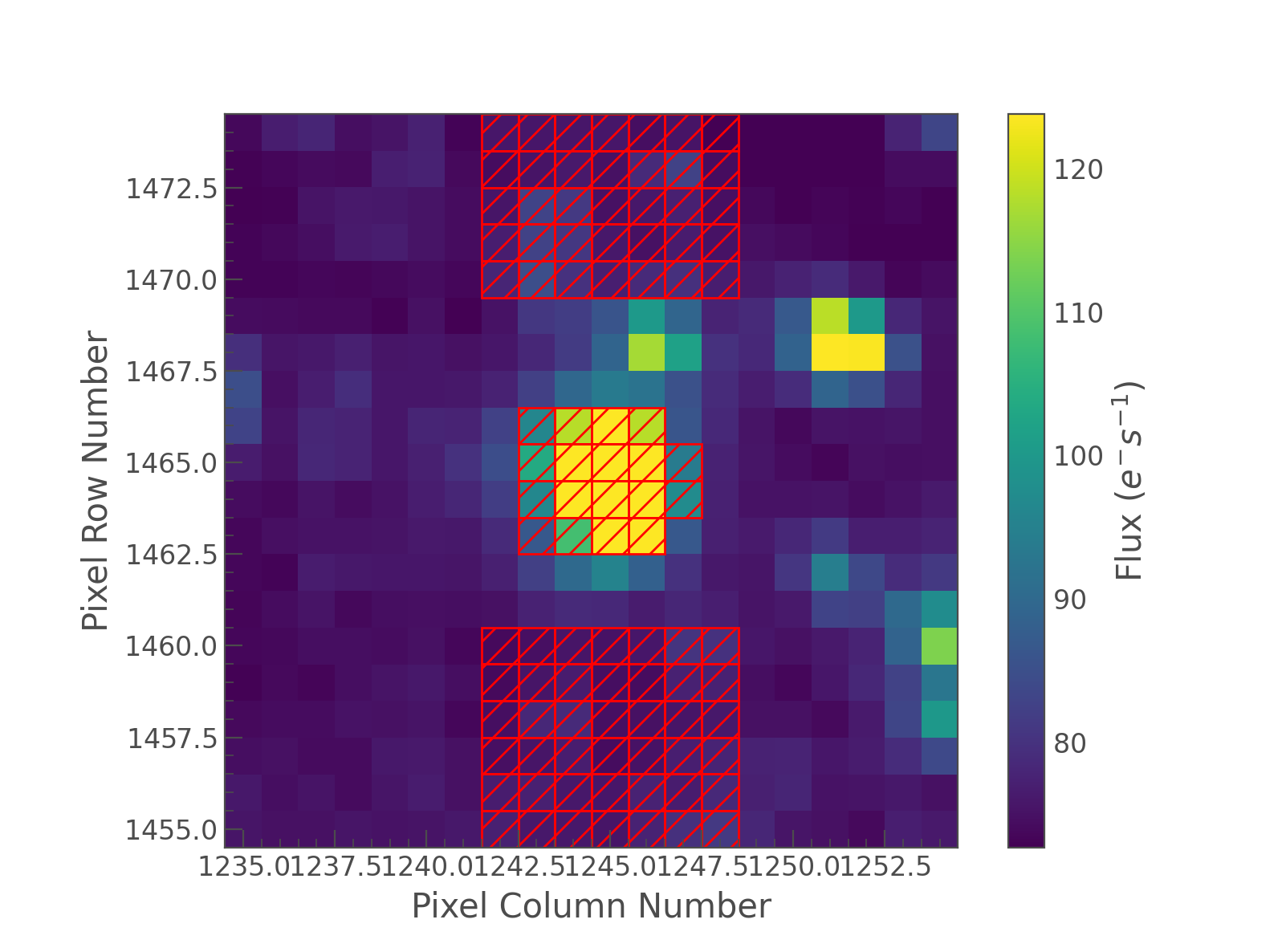}
  \includegraphics[width=0.45\textwidth]{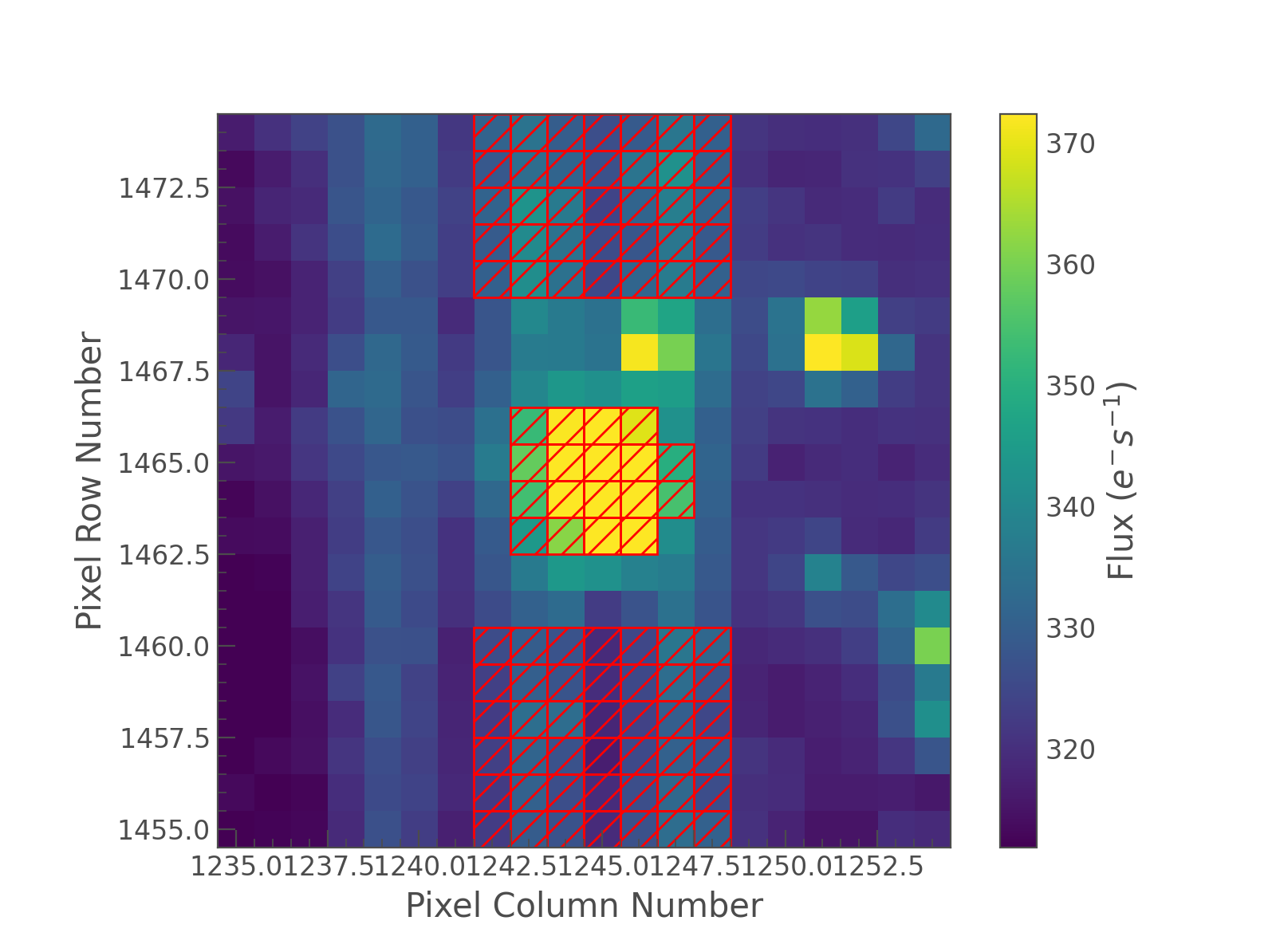}
  \includegraphics[width=0.45\textwidth]{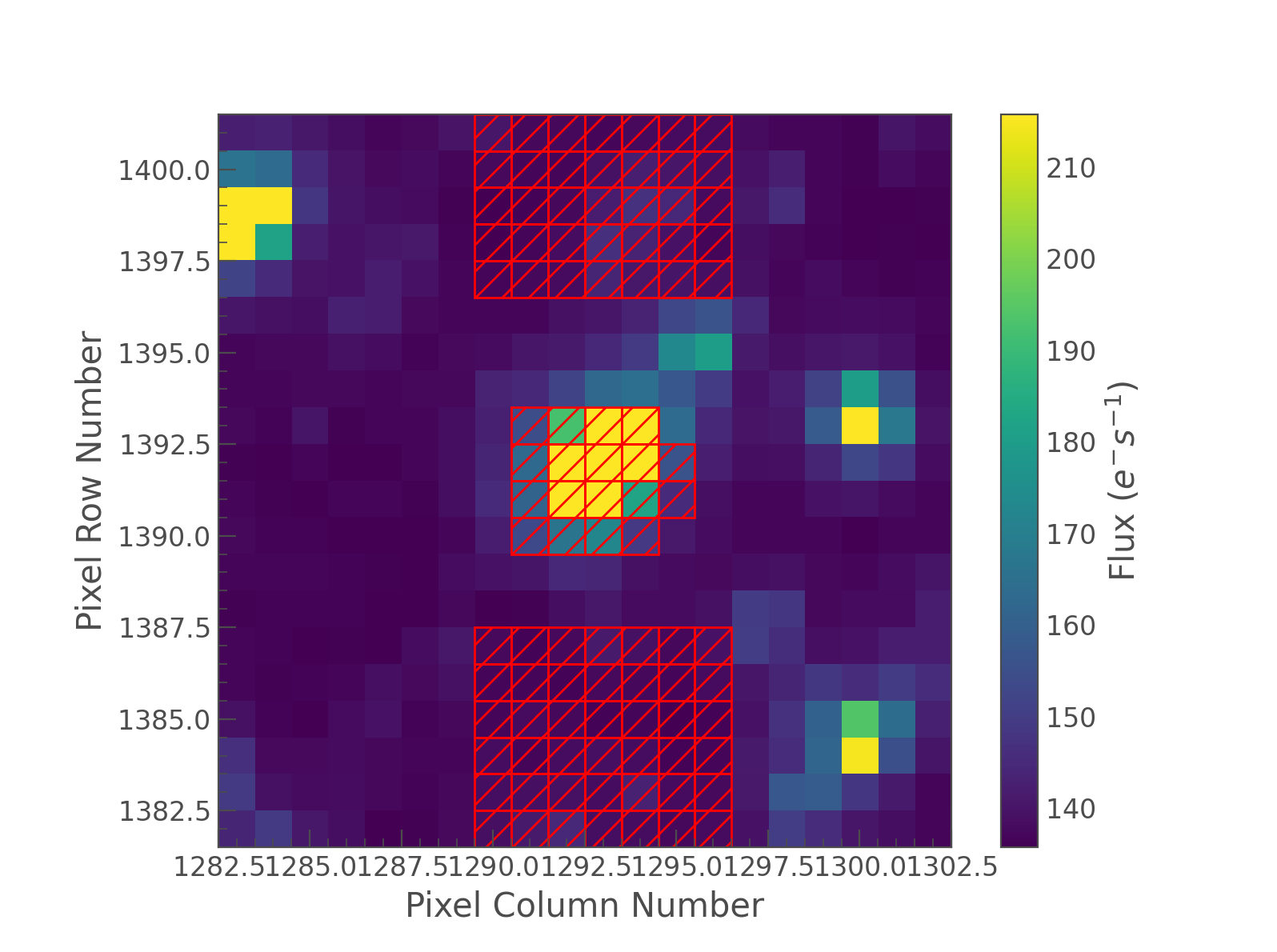}
  \includegraphics[width=0.45\textwidth]{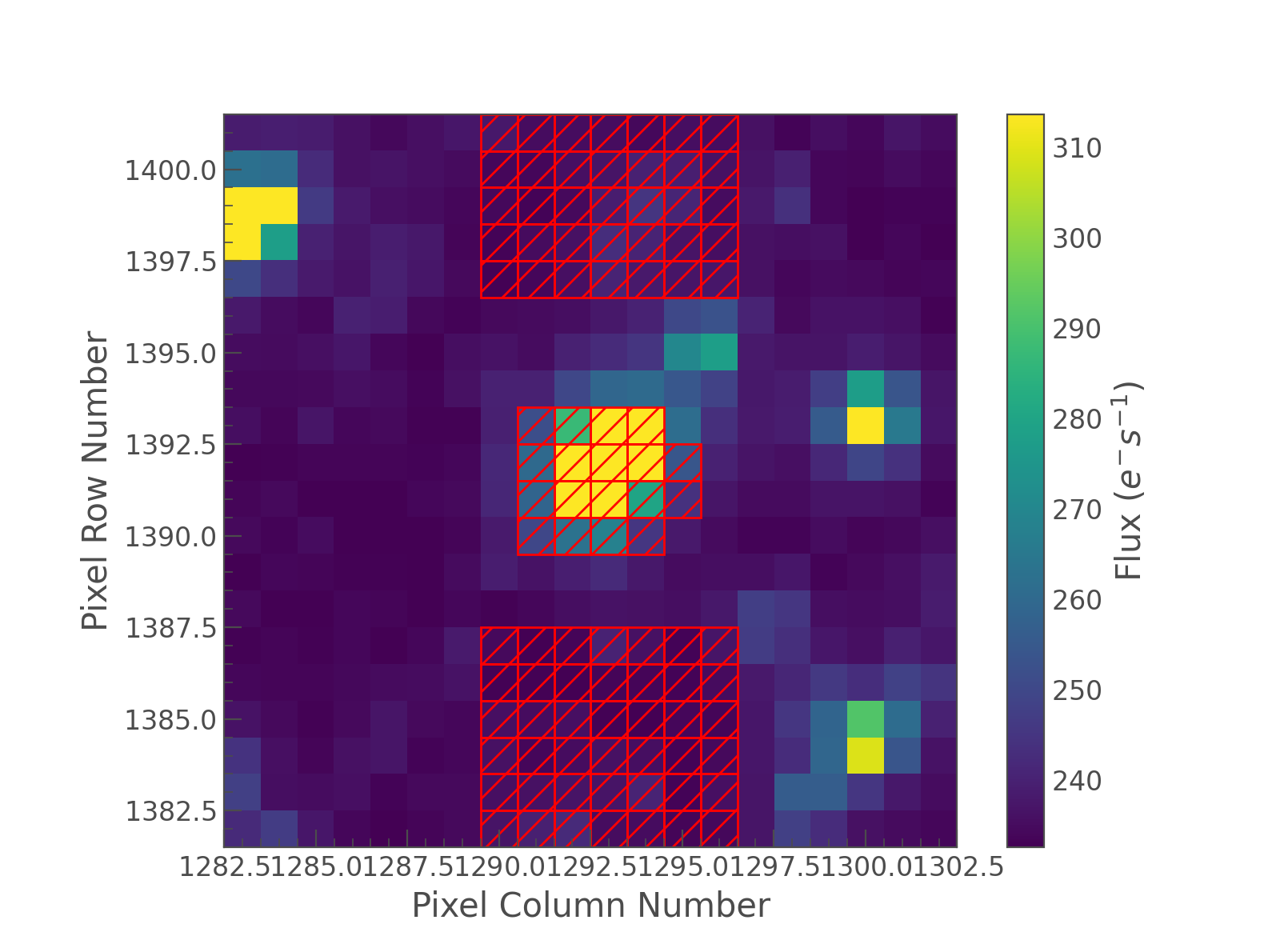}
  \caption{FFIs from Sector 50 (upper panels) and 51 (lower panels). 
  Apertures used for flux extraction are the areas marked in red 
  in the centers of the images. Areas used for background estimation 
  are the rectangular regions above and below the aperture. The 
  upper right panel is an example of an image from Sector 50 where 
  a banded structure can be seen in the background caused by scattered 
  light and the reflective metal straps on the backs of the CCDs.}
  \label{fig:bands}
\end{figure*}
An issue not encountered with the Sector 23 and 24 data was 
scattered light affecting the Full Frame Images (FFIs). As can 
be seen in Figure \ref{fig:bands}, part of the Sector 50 data suffered 
from visible bands as a background in the image. This is a well-known 
problem caused by the reflective metal straps on the backs of 
the CCDs \citep[see Section 6.6.1 of][]{tessinstrhb}. In Sector 
50 this banding structure only appeared in a subset of the images, 
but caused us to change our background subtraction process. Our 
solution was to use fixed regions of the images for the aperture and 
background, instead of using pixels above or below a certain threshold 
in the entire cutout image. The aperture used for flux extraction can 
be found in the centers of the images in Figure \ref{fig:bands}, 
and the areas used for background are the rectangular areas above 
and below the aperture. Sector 51 did not suffer from this banding 
effect, but we used the same definitions for the aperture and 
background for consistency.

Other than this change for the definitions of the aperture and 
background, the same procedures were used as in 
\citetalias{original}. The \texttt{ATARRI} package \citep{atarri} 
was used for an initial analysis, \verb|search_tesscut| 
\citep{tesscut} from the \verb|lightkurve| python package 
\citep{lightkurve} was used to download FFI data, and a frequency 
analysis was done using \verb|Period04| \citep{period04}.

\begin{figure*}[htb]
  \centering
  \includegraphics[width=0.99\textwidth]{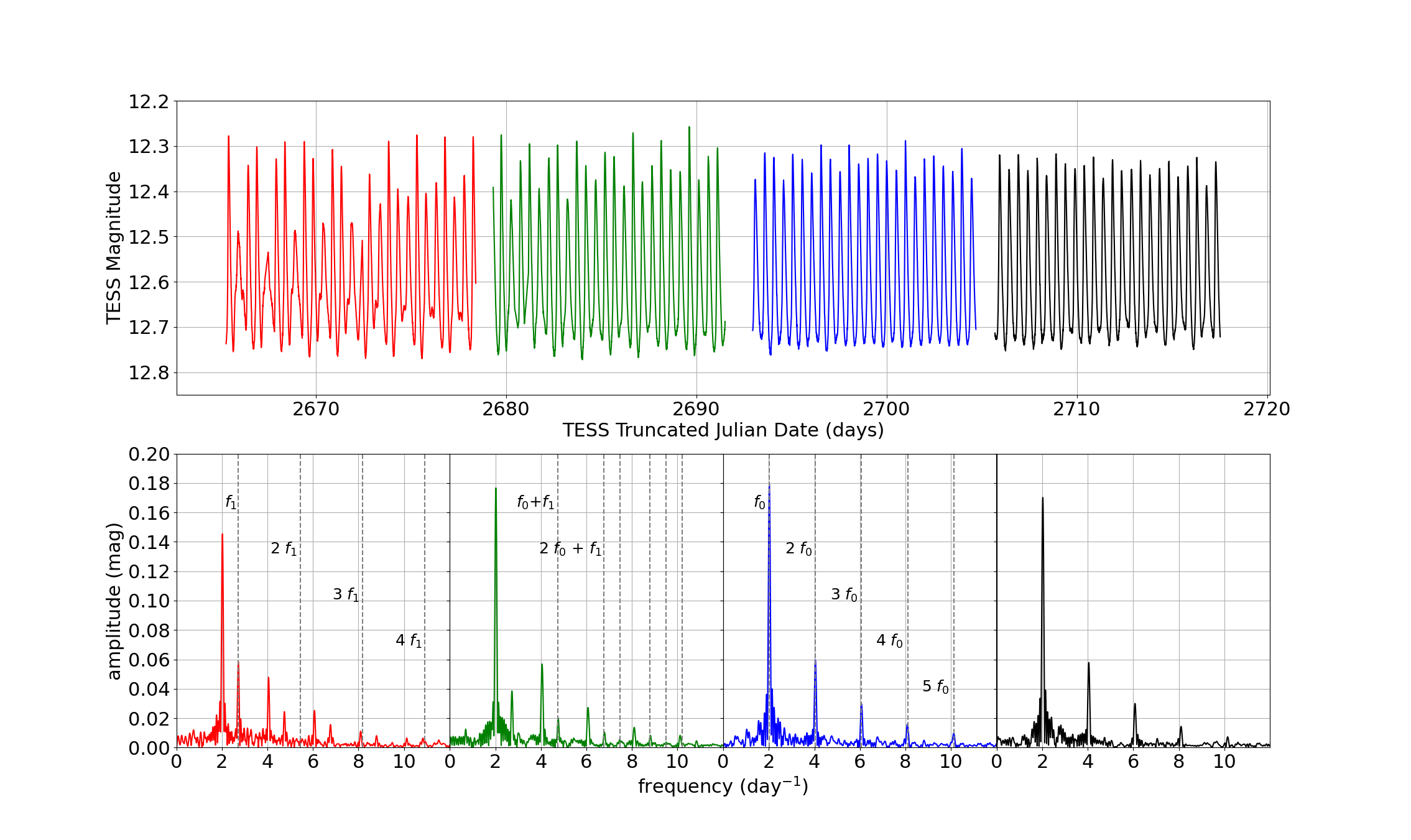}
  \caption{The upper panel shows the full light curve of V338 Boo 
  from Sectors 50 (red and green) and 51 (blue and black) from TESS 
  FFIs. The lower panels are a frequency analysis of the color-coded 
  segments from the upper plot. Dashed lines and text label the locations of 
  various peaks, with $f_0$ and $f_1$ corresponding to the fundamental 
  and first overtone frequencies, respectively.}
  \label{fig:lightcurve}
\end{figure*}
The extracted light curve from TESS Sectors 50 and 51 
can be found in Figure \ref{fig:lightcurve}. This figure is the same 
as Figure 1 from \citetalias{original} but for the newer sectors 
of data. A comparison between the figures shows that 
the last orbit of Sector 24 looks very similar to the first orbit 
of Sector 50 in both the light curve shape and the frequency 
analysis.

\subsection{AAVSO Follow-up Observations}\label{ssec:aavsodata}
In anticipation of the newer TESS results for V338 Boo, and in order 
to see the behavior of this star after the TESS observing window, an alert 
was issued through the AAVSO. Alert Notice 786 (issued 2022 July 12) asked 
the AAVSO community to observe this star so that we could extend the 
observations and analysis from TESS.

\begin{table}[htb]
  \centering
  \begin{tabular}{|l|r|r|}
  \hline
  Observer & \# of $V$ & \# of \\
  Code & Meas. & Nights \\
  \hline
  \hline
  PDM  &  1105  &  21  \\
  HGAG  &  851  &  8  \\
  BSM  &  387  &  7  \\
  SFV  &  267  &  7  \\
  NKEA  &  255  &  6  \\
  LDRB  &  374  &  5  \\
  ATE  &  1017  &  4  \\
  VTY  &  234  &  4  \\
  MFAA  &  170  &  4  \\
  SABB  &  159  &  4  \\
  RRO  &  95  &  4  \\
  \hline
  BMN  &  483  &  2  \\
  MZK  &  333  &  2  \\
  SAH  &  230  &  2  \\
  AANF  &  83  &  2  \\
  GALF  &  43  &  2  \\
  GCHB  &  80  &  1  \\
  CCHD  &  60  &  1  \\
  MMAO  &  7  &  1  \\
  BDQ  &  1  &  1  \\
  \hline
  \end{tabular}
  \caption{A list of the number of individual $V$ magnitude 
  measurements and number of different nights observed by 
  each of the AAVSO observers that contributed in the summer 
  of 2022 to AAVSO Alert Notice 786.}
  \label{tab:aavso}
\end{table}
In total, 9,478 individual brightness measurements were submitted by 
23 different observers between 2022 July 12 and 2022 August 14. Out of that 
total, there were 6,234 quality 
measurements (errors $<$0.025 mag) in the $V$ band by 20 different 
observers. These observers and their contributed number of measurements 
and observing nights are listed in Table \ref{tab:aavso}.

\section{Results}\label{sec:results}
\subsection{TESS}\label{ssec:tess}
As can be seen in Figure \ref{fig:lightcurve}, there is a definite change 
in both the light curve shape and frequency analysis of different portions 
of the light curve as observed by TESS in Sectors 50 and 51. In particular, 
in Sector 51 it appears that the first overtone mode pulsation has 
completely disappeared from the frequency analysis.

\begin{figure*}[htb]
  \centering
  \includegraphics[width=0.99\textwidth]{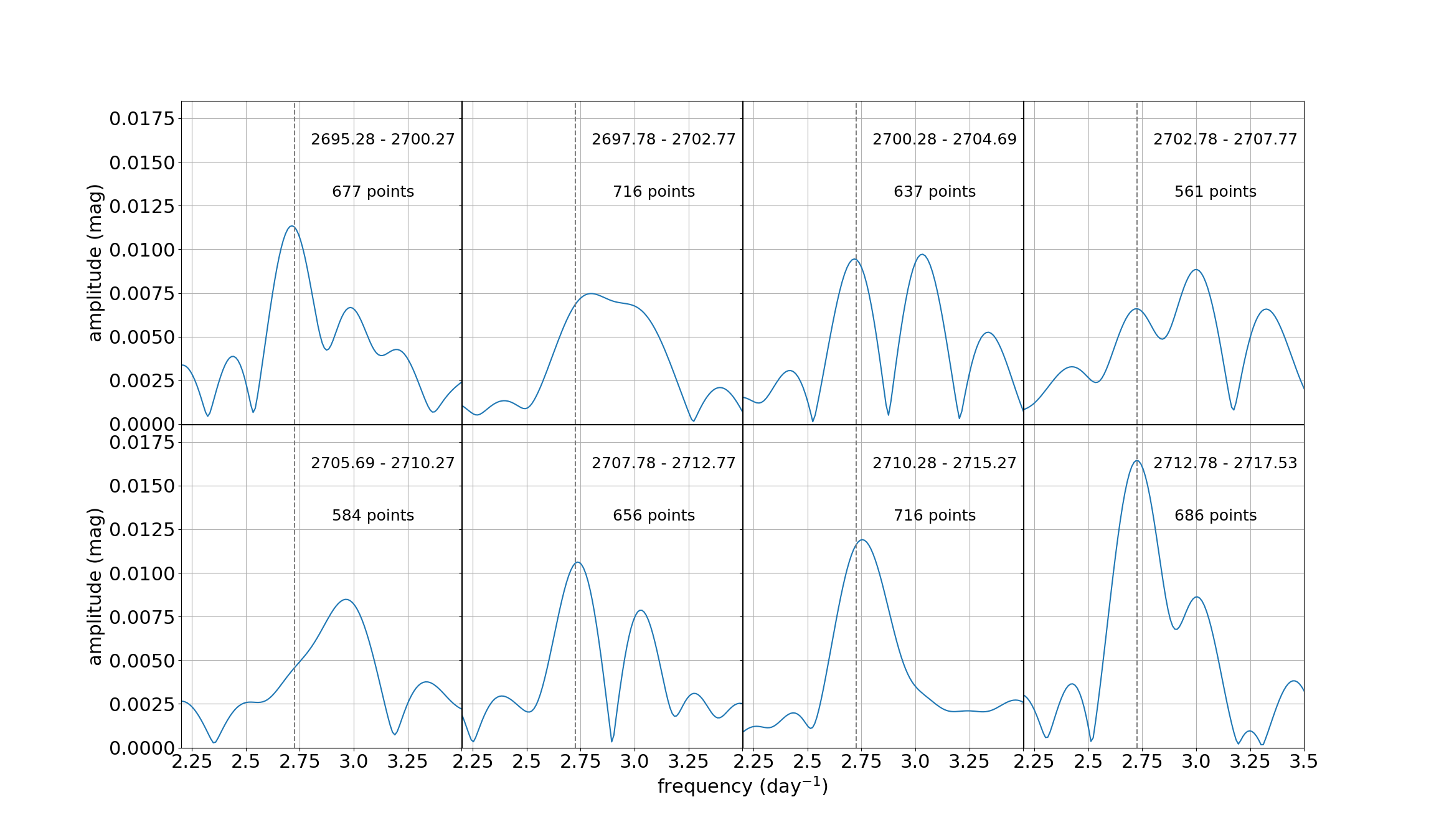}
  \caption{Frequency analysis of V338 Boo in the frequency 
  range around its first overtone mode. Each panel is a 5-day window 
  of data, increasing by 2.5 days at each step from left to right and 
  10 days from top to bottom. The TESS Julian date range is given in the upper 
  right of each panel, and below that is the number of data points in that 
  date range. The vertical dashed line shows the location of the first overtone 
  mode frequency.}
  \label{fig:min}
\end{figure*}

We examined Sector 51 data in more detail to see if the first overtone 
mode completely disappears, or just becomes very small. To do this, 
we analyzed 5-day segments of data and stepped forward by 
2.5 days. Figure \ref{fig:min} shows a timeframe around the 
smallest peak of the first overtone mode.
In the first 5-day window (upper left panel of Figure \ref{fig:min}), 
which corresponds to a TESS Julian date (TJD = JD - 2457000) window 
of 2695.28 - 2700.27, there is still evidence of the first overtone mode 
peak with an amplitude of about 11 mmag. Ten days later (lower left 
panel of Figure \ref{fig:min}) this pulsation mode is completely gone - 
there is no obvious peak in the frequency analysis. Subsequent 5-day 
windows see the first overtone mode reappear and grow in size. This 
means that for a brief period of time, only a few days at most, around a 
TJD of about 2707, V338 Boo was not pulsating in the first overtone 
mode.

In \citetalias{original} we saw an increasing fundamental mode 
and a decreasing first overtone mode in this star, but the TESS 
observations ended before the maximum (minimum) in the 
change was reached. In this data, however, we cross that extreme 
point in the transient behavior and see that the first overtone 
mode does not just reach some minimum pulsation amplitude, 
it disappears altogether. Although there has been previous evidence of 
stars switching their pulsation modes from RRd to RRab or 
vice versa \citep[see e.g.~][]{v79,v79back,ogle1,ogle2,ogle3,css},
this is the first time this process has been seen in real time.

\begin{figure*}[htb]
  \centering
  \includegraphics[width=0.99\textwidth]{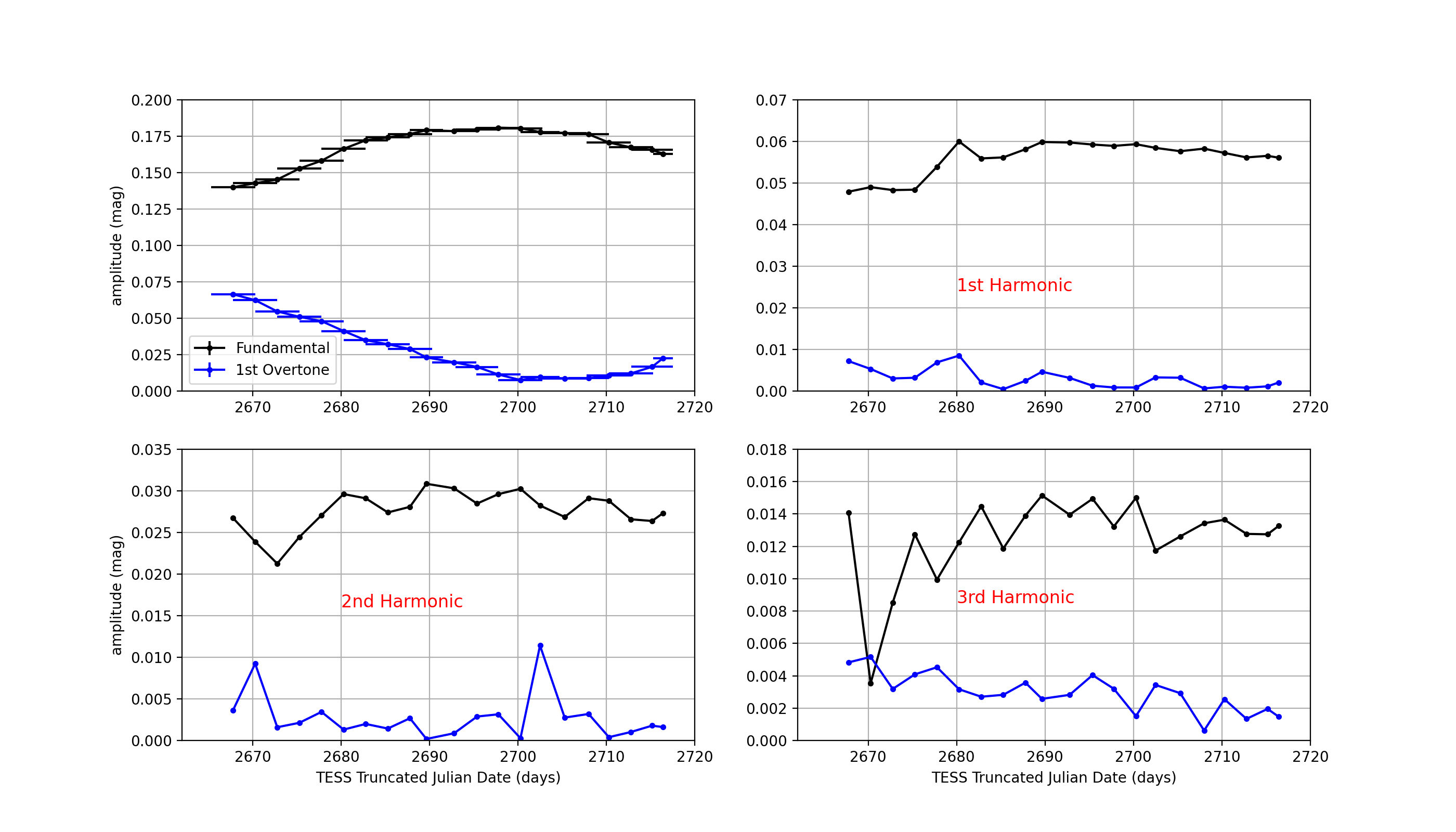}
  \caption{Amplitude of the fundamental mode (black) and 
  first overtone mode (blue) for the primary peak (top-left panel), 
  first harmonic (top-right), second harmonic (bottom-left), 
  and third harmonic (bottom-right) for V338 Boo. This is the 
  same as Figure 3 of \citetalias{original} but with data from 
  TESS Sectors 50 and 51.}
  \label{fig:trend}
\end{figure*}
In Figure \ref{fig:trend} we show the amplitude of the peaks 
in the frequency analysis as a function of time. This procedure 
is the same as in \citetalias{original} with the new data - we 
analyze 5-day windows, shifting by 2.5 days through both 
sectors of data. While in \citetalias{original} there seemed 
to be a monotonic increase in the fundamental mode 
amplitude, in Figure \ref{fig:trend} we see that this amplitude 
reaches a peak at a TJD of about 2695 and then begins 
to decrease.

Similarly, we see that the first overtone mode in the newer 
data decreases, reaching a minimum value before starting 
to increase. However, as discussed above, this minimum 
is reached approximately 10 days later. In the upper left 
panel of Figure \ref{fig:trend} the first overtone mode 
amplitude never reaches exactly zero because we find the 
largest amplitude in a narrow frequency range centered on 
the known pulsation frequencies, so at the minimum we are 
measuring noise in the analysis.

\begin{figure*}[htb]
  \centering
  \includegraphics[width=0.99\textwidth]{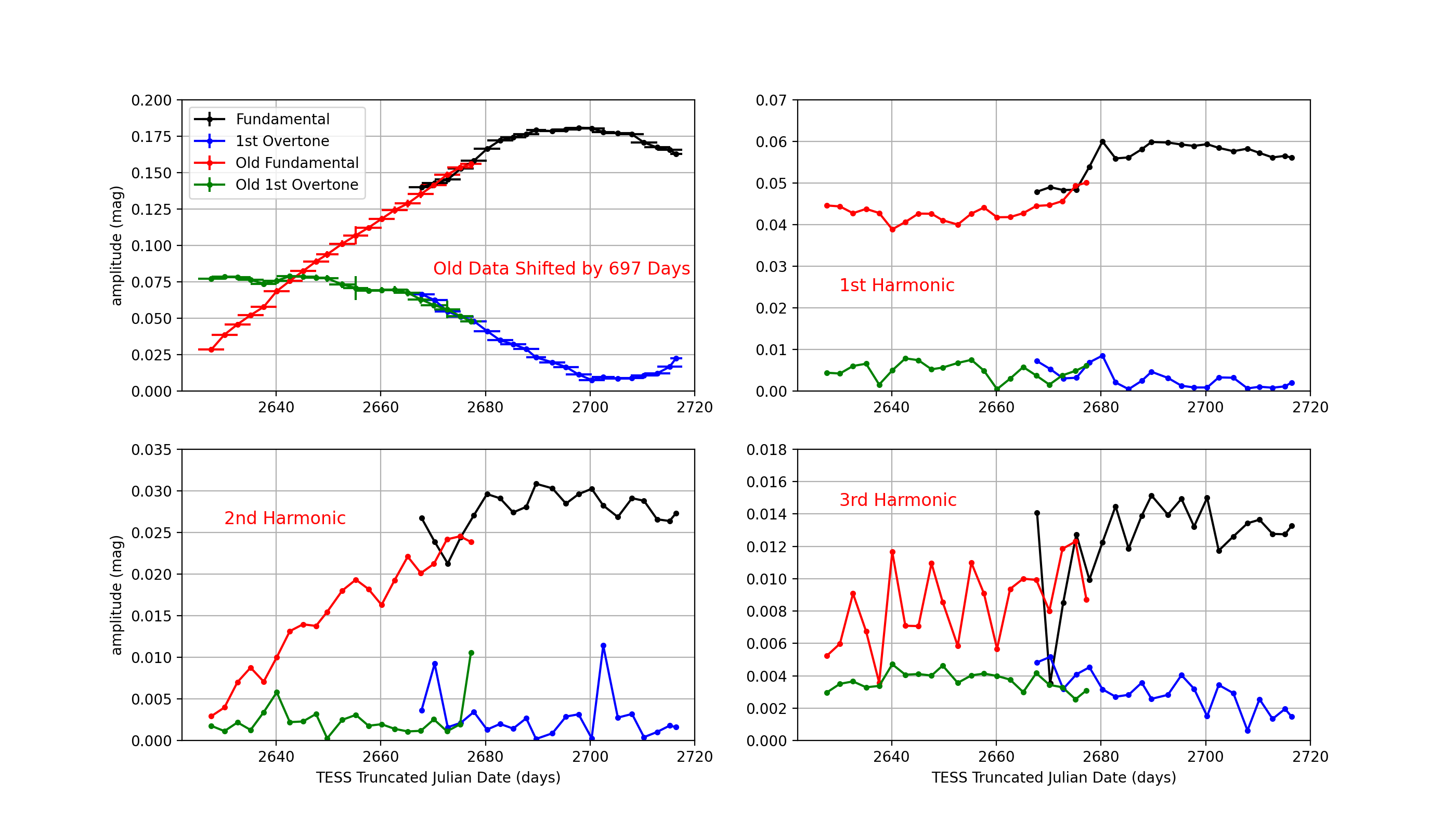}
  \caption{The same as Figure \ref{fig:trend} with the data 
  from Sectors 23 and 24 \citepalias{original} shifted in time 
  by 697 days and overlaid.}
  \label{fig:trendwithold}
\end{figure*}
By comparing Figures 1 and 3 of \citetalias{original} with 
Figures \ref{fig:lightcurve} and \ref{fig:trend} here, it seems 
as if the behavior seen in V338 Boo is not just transient in 
nature, but is periodic. The light curve shape and 
frequency analysis of the second orbit of Sector 24 
looks very similar to the first orbit 
of Sector 50. The second orbit of Sector 24 started on 
2020 February 29, and the first orbit of Sector 50 started 
on 2022 March 26, which is a difference of 697 days 
(inclusive). In Figure \ref{fig:trendwithold} we show the 
trends in the amplitudes of the different modes of Sectors 
50 and 51, along with Sectors 23 and 24 shifted by 697 
days.

There is excellent agreement between the amplitudes 
in Sectors 24 and 50 for the main pulsation modes, but 
also for their harmonics. Additionally, the two trends match 
up extremely well. The fact that the exact difference between 
the starts of Orbit 2 of Sector 24 and Orbit 1 of Sector 50 give a 
shift in time that matches the periodic behavior of V338 Boo is 
purely coincidental.

\subsection{AAVSO}\label{ssec:aavso}
As mentioned in Section \ref{ssec:aavsodata}, Alert Notice 
786 was sent out to the AAVSO community on 2022 July 12, 
and the first observations for this alert were taken that night. 
These observations were critical because the TESS observations 
showed that the first overtone mode completely disappeared 
for a brief period, but was again seen at the end of Sector 51. It 
was unknown, however, if the first overtone mode continued to 
increase in amplitude, or if it remained very small. 
The last day of observations of Sector 51 was 2022 May 6, so 
the AAVSO observing campaign began two months after this.

\begin{figure*}[htb]
  \centering
  \includegraphics[width=0.99\textwidth]{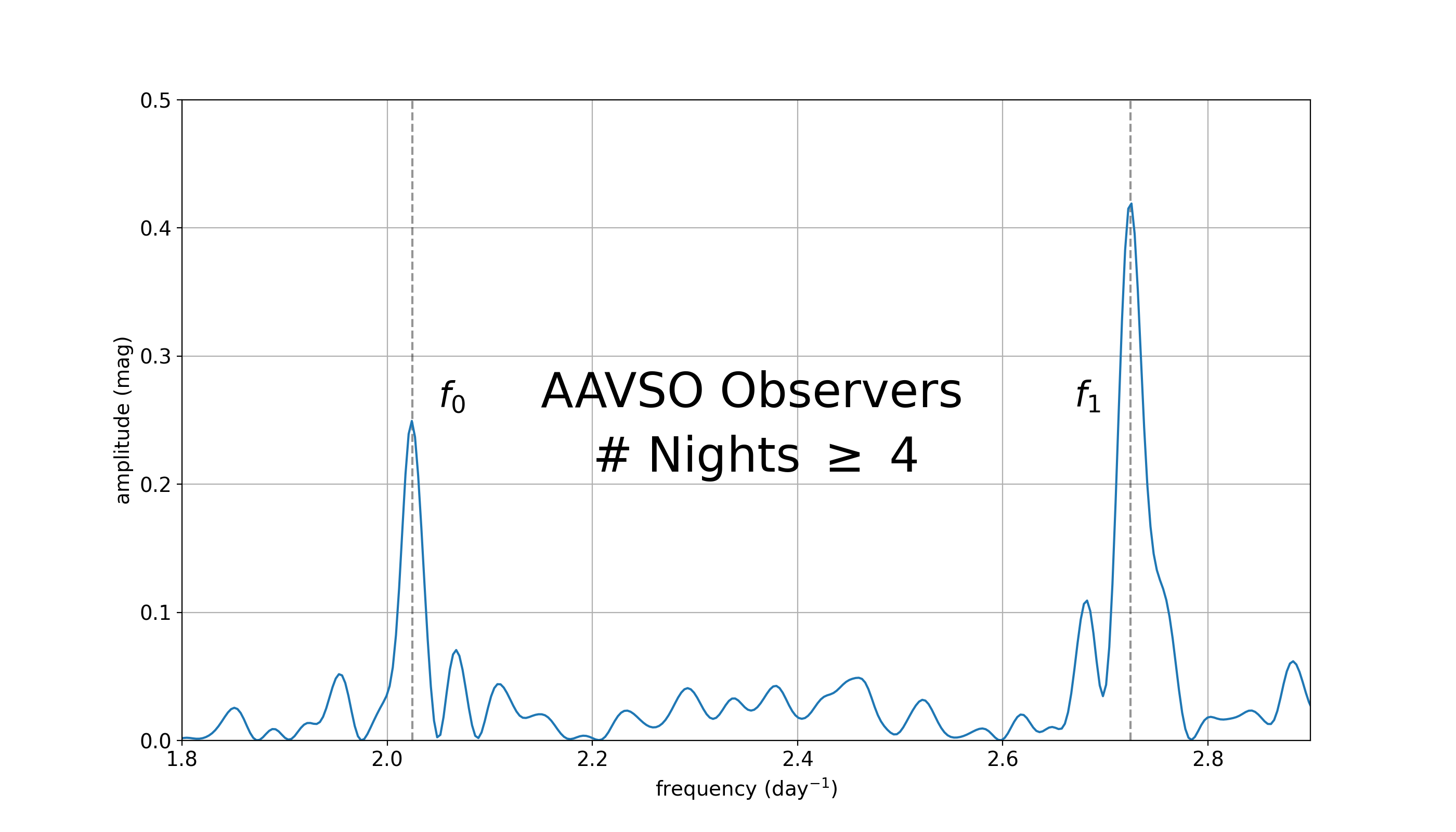}
  \caption{Frequency analysis of observations taken for Alert Notice 786 
  of the AAVSO by observers with 4 or more nights of observations. The 
  fundamental mode frequency ($f_0$) and first overtone mode 
  frequency ($f_1$) are shown with dashed vertical lines.}
  \label{fig:aavso}
\end{figure*}
There were 11 AAVSO observers who took brightness measurements 
in the $V$ band of V338 Boo for Alert Notice 786 on 4 or more nights during 
the campaign. Importantly, observer PDM submitted over 1,100 observations 
on 21 different nights, which served as a comparison for several of the other 
data sets. We combined data from observers with 4 or more nights of 
observations to do a frequency analysis of V338 Boo over 34 days in the summer 
of 2022, which is similar in length to a single TESS sector. The observations 
of HGAG were shifted by +0.24 mag and the observations of ATE were shifted 
by +0.04 mag to match the observations of the other observers. The results 
of the frequency analysis are shown in Figure \ref{fig:aavso}. During 
this timeframe, the first overtone mode pulsation is again dominant, 
similar to the beginning of Sector 23 \citepalias{original}.

\begin{figure*}[htb]
  \centering
  \includegraphics[width=0.99\textwidth]{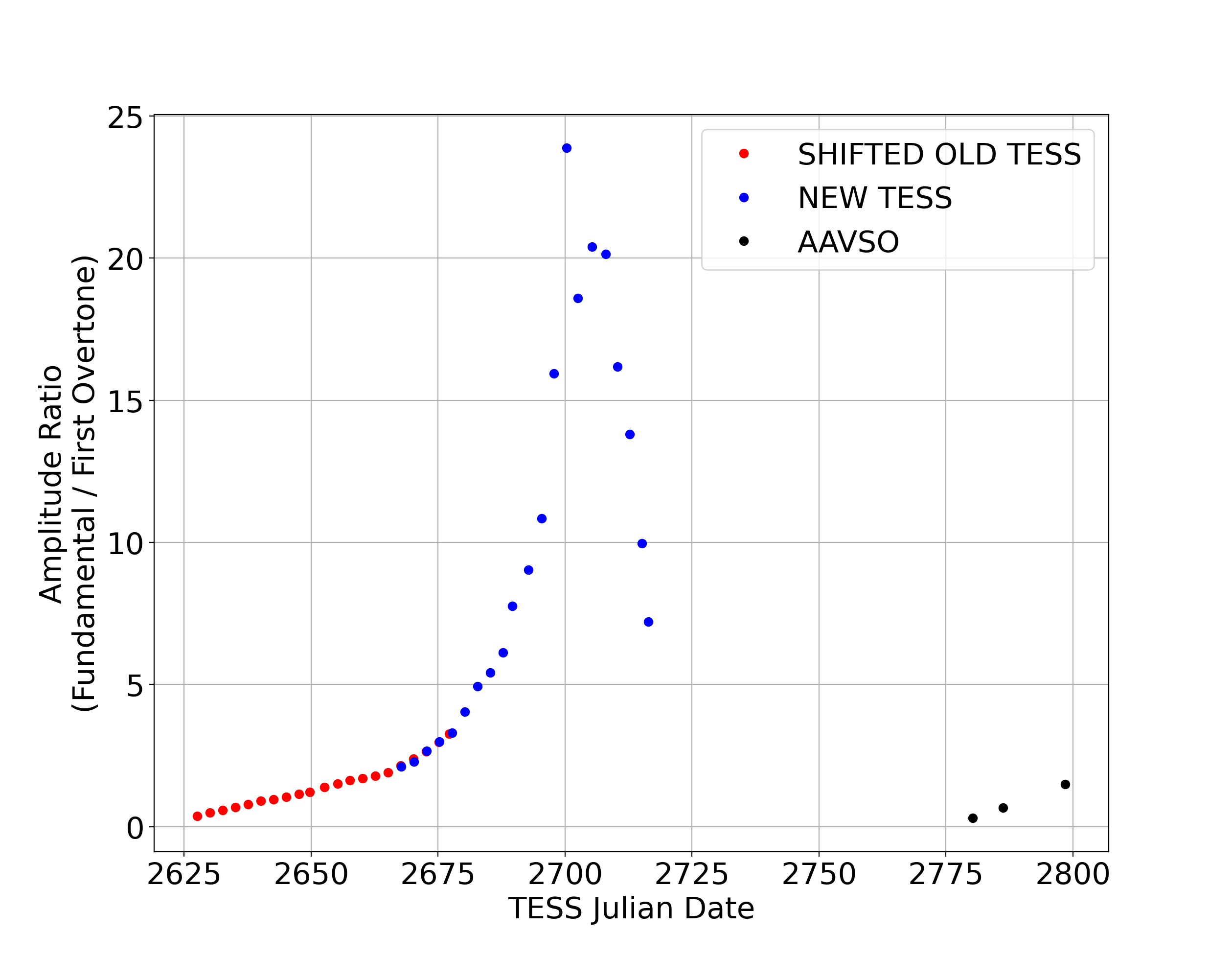}
  \caption{The amplitude ratios of the fundamental mode pulsations 
  to the first overtone mode pulsations for TESS data from 2020 
  (red points) and 2022 (blue points), and the AAVSO observing 
  campaign (black points).}
  \label{fig:amprats}
\end{figure*}
Because of the amount and duration of the AAVSO campaign observations, 
we were also able to split the sample up to look for trends in the 
amplitude ratios of the fundamental to first overtone modes. We 
looked at three ``halves'' of the data - the first half, the second half, 
and the middle half (half the observing window centered in the middle 
of the observations). While the overall AAVSO data shows that the 
first overtone mode is dominant (see Figure \ref{fig:aavso}), we also 
see in Figure \ref{fig:amprats} that the amplitudes of the two pulsation 
modes are changing in that data set as well.

\section{Conclusions and Discussion}\label{sec:conclusions}
The observations from TESS and the AAVSO campaign show us that 
the changes initially seen in V338 Boo appear to be periodic in nature 
and not just transient. This is an important distinction, and is 
helpful in determining the possible physical mechanisms 
causing this behavior.

Additionally, the first overtone mode completely 
disappears for a brief period of time, and then reappears 
shortly after. Therefore, either something in the stellar interior 
is regenerating this pulsation mode, or there is a ``memory'' 
in the star that restarts this mode \citep[e.g.~][]{stothers}.

\subsection{Mode-Switching Behavior}\label{ssec:modeswitch}
There are still unanswered questions related to this object. First, 
is this ``mode-switching'' behavior? Previous results 
\citep[see e.g.~][]{v79,v79back,ogle1,ogle2,ogle3,css} have shown 
that other RR Lyrae have been observed as one type (e.g.~RRd) and 
in subsequent observations they are a different type 
(e.g.~RRab).

In \citet{v79M3} they examine the behavior of the 
RR Lyrae variable V79 in the M3 globular cluster over $\sim$50 years 
of observations. Many of the years had very few observations, and 
therefore the authors restricted some of the analysis to only those 
years with more than 40 observations. This resulted in 11 years from 
the historical data in which they determined the primary period of 
pulsation for this star. They saw there was a large variation in the primary 
period - both in the value of the period itself, and in the primary 
mode in which the star appeared to be pulsating. In the year 1976 there were 
108 observations, by far the most of any year in the historical data, and 
this allowed the authors to look for changes on smaller timescales. 
In their Figure 2, there are changes apparent on the timescale 
of approximately one month.

This analysis and the results are 
very similar to what we saw in \citetalias{original} with previous data 
of this star 
\citep{oaster06,jaavso,jaavso2} - the sparse sampling of the 
periodic changes showed that V338 Boo was behaving strangely, 
but it was unclear why. So, it is possible, if not likely, that RR Lyrae 
dubbed as ``mode-switching'' from sparsely sampled data are in 
fact exhibiting periodic changes just like those seen in V338 Boo.

\subsection{Blazhko Effect}\label{ssec:blazhko}
A second question: is this related to the Blazhko effect seen most 
commonly in RRab stars, but also seen in RRc stars? As shown here, 
the changes of V338 Boo seem to be periodic, and if this star were 
an RRab or RRc, it would certainly be considered a Blazhko 
variable.

A useful comparison can be found in \citet{RRdM3}. 
They identified four RRd stars in the globular cluster M3 that 
exhibited modulations of both the fundamental and first overtone 
mode pulsation amplitudes, which they attributed to the Blazhko 
effect. There are some important differences 
between those stars and V338 Boo, however.

First, for both the M3 stars and V338 Boo there are periodic amplitude 
modulations of both pulsation modes. But as shown here, there is a 
short period where the first overtone mode completely disappears from 
V338 Boo (we note that this could not be seen in \citet{RRdM3} even if 
it were present because of the sparse sampling of the light curves).

Second, all four stars in \citet{RRdM3} had period ratios of 
the radial pulsation modes that were anomalous. 
However, the period ratio of the pulsation modes of V338 Boo is well 
within what is considered normal for RRd variables.

And finally, only one of the four RRd stars in \citet{RRdM3} had 
amplitude changes of the modes that were synchronized in any way 
(anti-correlated in this case). The periodic changes we see in amplitudes 
of the two pulsation modes of V338 Boo are anti-correlated, which matches 
the one example in \citet{RRdM3}.

An anti-correlation is an important fact when trying to determine the 
cause of these changes. This likely means that the changes seen 
in V338 Boo are coupled, or drawing from and storing to the same 
energy source. These changes could be due, for example, to competition 
in the $\kappa$ mechanism itself \citep{mos15}, or changes in 
turbulent motions that generate transient magnetic fields 
\citep{stothers}.

We also note that three of the M3 RRd stars from \citet{RRdM3} 
had amplitude changes that were not synchronized, and even 
the periods of the amplitude changes within the same star were 
different. This could mean that stars labelled as ``mode-switching'' 
are actually changing periodically in the same manner - when 
they are observed as RRab stars, the first overtone mode is in a 
long period of very low amplitude. Mode-switching, therefore, 
may be part of a periodic change in the pulsations and not an 
abrupt, transient phenomenon. In fact, as shown in \citet{lsher}, 
transient behavior can be mimicked by the constructive and 
destructive interference of periodic signals within RR Lyrae stars.

The term ``Blazhko Effect'' is loosely defined to encompass 
many different types of changes in the light curves of RR Lyrae. 
As such, we argue that V338 Boo should be considered a 
variable of this type. But is the same physical mechanism causing 
the changes seen in V338 Boo and those seen by \citet{RRdM3}? 
And more broadly in the RRab and RRc variables? Could behavior 
that seems to be abrupt and/or transient \citep[e.g.~][]{leborgne} 
actually be due to the interference of periodic signals 
\citep[e.g.~][]{lsher}? The Blazhko effect seems to occur much 
more frequently in RRab types than RRc types (see, e.g., 
\citet{ogleab} and \citet{oglec} for incidence rates from 
the Optical Gravitational Lensing Experiment, and 
\citet{keplerab} and \citet{keplerc} for incidence rates 
from Kepler), but why is this the case? In other words, what 
physical conditions within RRab stars make them more likely 
to exhibit this phenomenon?

The mystery of the Blazhko effect may be best approached 
by studying double-mode RR Lyrae. V338 Boo exhibits 
periodic changes in both pulsation modes, and those changes 
are anti-correlated. However, this is not the case with all 
Blazhko RRd stars \citep{RRdM3,smo15}. Identifying 
commonalities and differences between these RRd stars 
will help point us to physical properties that could be 
driving the changes. It may also mean that different 
physical mechanisms are found to be responsible for 
behavior currently labelled as the Blazhko effect.

\section*{Acknowledgements}
K.C. would like to acknowledge funding from the National 
Science Foundation LEAPS-MPS program through Award 
\#2137787.

In addition to the AAVSO observers listed as co-authors, 
we would like to acknowledge the contributions of the 
following observers for contributing to the Alert 
Notice 786 campaign: Ronald Blake (BMN), Kenneth 
Menzies (MZK), Gerard Samolyk (SAH), Antonio 
Agudo (AANF), Alberto Garcia Sanchez (GALF), 
Charles Galdies (GCHB), Charles Cynamon (CCHD), 
Mario Morales Aimar (MMAO), and Alan Bedard (BDQ).

\bibliography{paper}{}
\bibliographystyle{aasjournal}

\end{document}